\documentclass[preprint,journal]{vgtc}                     

\onlineid{1116}

\ieeedoi{10.1109/TVCG.2023.3327172}

\vgtccategory{Research}

\vgtcpapertype{Representations \& Interaction}

\title{DIVI: Dynamically Interactive Visualization}

\author{%
  \authororcid{Luke S.\ Snyder}{0000-0002-4340-8263} and
  \authororcid{Jeffrey Heer}{0000-0002-6175-1655}
}

\authorfooter{
  \item
  	Luke S. Snyder and Jeffrey Heer are with the University of Washington.
  	E-mail: snyderl, jheer@cs.washington.edu

}

\abstract{%
  Dynamically Interactive Visualization (DIVI) is a novel approach for orchestrating interactions within and across static visualizations.
DIVI deconstructs Scalable Vector Graphics charts at runtime to infer content and coordinate user input, decoupling interaction from specification logic. 
This decoupling allows interactions to extend and compose freely across different tools, chart types, and analysis goals.
DIVI exploits positional relations of marks to detect chart components such as axes and legends, reconstruct scales and view encodings, and infer data fields.
DIVI then enumerates candidate transformations across inferred data to perform linking between views.
\textcolor{red}{To support dynamic interaction without prior specification,} we introduce a taxonomy that formalizes the space of standard interactions by chart element, interaction type, and input event.
\textcolor{red}{We demonstrate DIVI's usefulness for rapid data exploration and analysis through a usability study with 13 participants and a diverse gallery of dynamically interactive visualizations, including single chart, multi-view, and cross-tool configurations.} 

}


\teaser{
    \includegraphics[width=\linewidth]{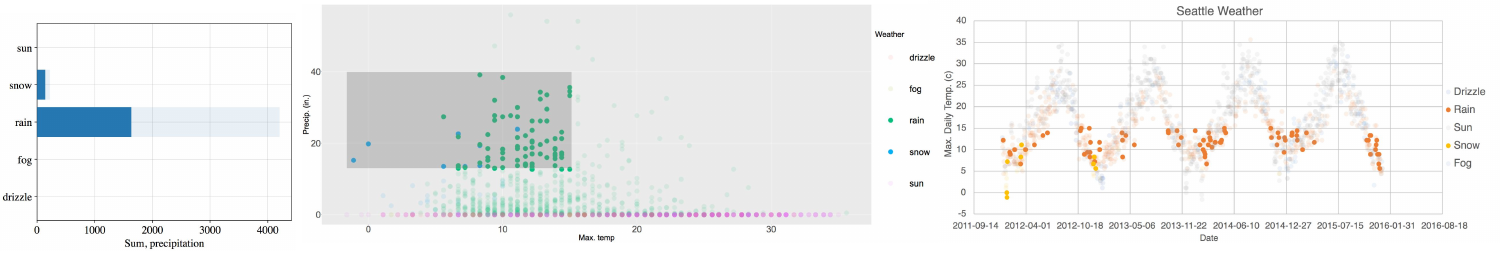}
    \vspace{-15pt}
    \caption{\textcolor{red}{Automatic multi-view, cross-tool interactions with DIVI. Dynamically interactive charts of Seattle weather data, from left to right: Matplotlib bar chart (\texttt{weather} vs. \texttt{sum(precip)}), ggplot2 scatter plot (\texttt{temp\_max} vs. \texttt{precip}), and Excel scatter plot (\texttt{date} vs. \texttt{temp\_max}). DIVI automatically deconstructs charts to identify semantic components and coordinate user input.} Here DIVI links \texttt{temp\_max} and \texttt{weather} between scatter plots, and \textcolor{red}{also leverages the source dataset} to infer bar chart aggregation of \texttt{sum(precip)} over \texttt{weather}. Interactions are linked: the user brushes an area in the ggplot2 scatter plot with low \texttt{temp\_max} and high \texttt{precip}, re-aggregating the bar chart and propagating the selection to the Excel scatter plot to indicate seasonal wintry periods.
    }
    \label{fig:teaser}
}

\graphicspath{{figs/}{figures/}{pictures/}{images/}{./}} 

\usepackage[svgnames]{xcolor}
\usepackage{listings}
\usepackage{multirow}
\usepackage{fontawesome}
\usepackage{makecell}
\usepackage{graphicx}
\usepackage{booktabs}
\usepackage{dblfloatfix}
\usepackage{amsmath}
\usepackage{algorithm2e}

\definecolor{figred}{HTML}{e15759}
\definecolor{figblue}{HTML}{4e79a7}
\definecolor{figorange}{HTML}{f28e2c}
\definecolor{figturqoise}{HTML}{76b7b2}
\definecolor{red}{RGB}{0, 0, 0}

\begin{document}


\maketitle

\section{Introduction}

A variety of people including data analysts, journalists, and teachers use visualizations to explore, analyze, and communicate data.
Though a range of tools support visualization creation, authoring \emph{interactive} visualizations is notably difficult.
Visualization developers may implement interactions by modifying visual encodings using low-level event handlers (as in D3~\cite{bostock2011d3}) or declarative selection specifications (as in Vega-Lite~\cite{satyanarayan2016vega}). 
Graphical tools such as Lyra 2~\cite{zong2020lyra} support custom interaction design without writing textual code, while libraries such as ggplot2~\cite{ggplot2} and Matplotlib~\cite{matplotlib} support only static plots.

A central concern is that interactions are tightly coupled with chart specification logic.
For instance, to select or zoom within a D3 scatter plot, one must manipulate the underlying Document Object Model (DOM) representation, while the same interactions in Vega-Lite instead require declarative specifiers.
These interactions require increasingly complex modifications and state handling for linked charts.
As a result, interaction reuse and flexibility is limited across different tools and chart configurations needed for rapid exploratory analysis.
Evolving user preferences, data sources, and analysis goals may additionally bring alternative unforeseen interaction needs.

\begin{figure}[!t]
    \centering
    \resizebox{1\linewidth}{!}{\includegraphics{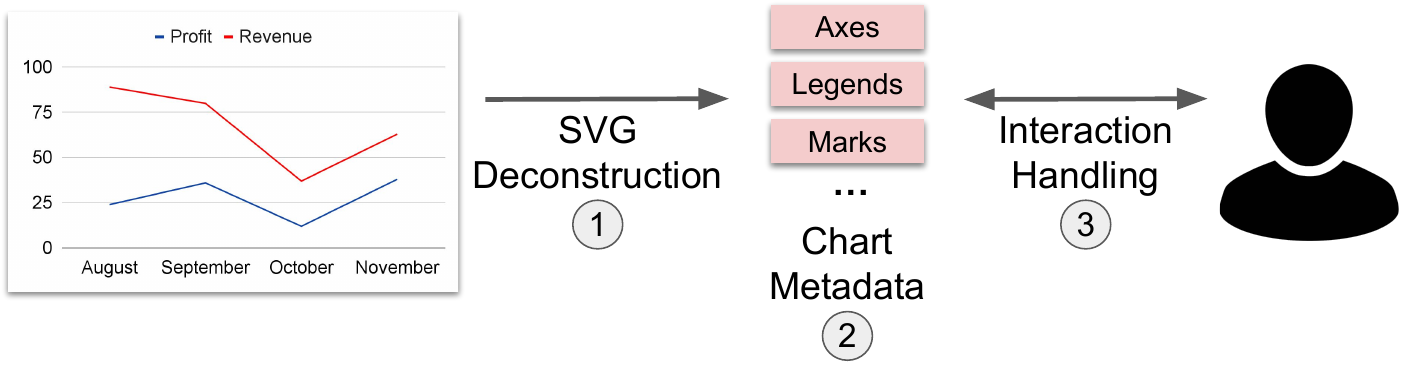}}
    \vspace{-15pt}
    \caption{\textcolor{red}{Overview of DIVI. DIVI enables dynamic interaction via a three-part process: (1) SVG deconstruction to automatically extract chart semantics; (2) standardized chart metadata to decouple specification and interaction logic, such as axes, legends, and marks; (3) automatic interaction handling to update chart structures based on user input.}}
    \label{fig:divi-overview}
    \vspace{-3mm}
\end{figure}

\textcolor{red}{
Informed by discussions with data analysts, we present \textbf{Dynamically Interactive Visualization (DIVI)}, a novel approach for automatically orchestrating interactions with static visualizations.
DIVI contributes a three-part model to enable dynamic interaction (Fig.~\ref{fig:divi-overview}): (1) deconstruct Scalable Vector Graphics (SVG) to automatically infer chart metadata, (2) leverage a standardized representation of chart semantics to decouple visual encoding specification and interaction logic, and (3) automatically handle user input and provide disambiguation support.
}

\textcolor{red}{
DIVI enables dynamic interaction rather than prior specification, allowing users to interact with charts in lieu of writing often complex and tedious interaction handling code.
We introduce a taxonomy that formalizes the space of standard user interactions by target visual element, interaction type, and input event to support dynamic orchestration and disambiguate user inputs (e.g., dragging to brush vs. pan).
We also delineate the chart metadata needed to perform each interaction and generate a standardized representation.
This standardization decouples the visual specification and interaction layers, allowing interactions to extend and compose freely across visualizations, including novel support for static tools (e.g., ggplot2~\cite{ggplot2} and Matplotlib~\cite{matplotlib}) as well as dynamic multi-view and cross-tool linking.
}

\textcolor{red}{
While it is theoretically possible to update popular visualization tools to expose sufficient chart metadata to enable dynamic interaction, it is infeasible in the short term given the internal differences among tools and the engineering work required.
DIVI instead leverages rendered SVG as a shared representation produced by a wide range of tools, deconstructing charts to infer the semantic components required for interaction.
The SVG format is sufficiently rich to infer chart semantics and can be modified directly to achieve interactive updates.
DIVI infers core chart structures---axes, legends, and marks---to reconstruct scales, data fields, and aspects such as chart orientation, stacking, and binning.
Meanwhile, access to underlying source data---either inferred from non-aggregated plots or provided directly---enables linked interactions across views.
DIVI enumerates candidate transformations across visualization pairs, using inferred data to detect potential links via attribute projection, aggregate transforms, and filtering.
}


\textcolor{red}{
To assess DIVI, we present a gallery of dynamically interactive charts and a usability study in which participants perform a variety of dynamic interactions and disambiguate their intents.
We also discuss SVG inference limitations and how users may correct them.
}
 

\textcolor{red}{
Our primary research contributions are:
\begin{enumerate}
    \item DIVI, a novel approach for dynamically orchestrating interactions in static visualizations, driven by a three-part model: (1) SVG deconstruction to infer chart semantics; (2) standardized chart metadata to decouple interaction from source tools; and (3) automatic interaction handling with disambiguation support. 
    \item A taxonomy to formalize user interactions by target visual element, interaction type, input event, and requisite chart metadata.
    \item SVG deconstruction techniques to infer necessary chart semantics and multi-view linkages.
    \item An example gallery of dynamically interactive visualizations and a usability study with 13 participants.
\end{enumerate}
}

\section{Related Work}
\label{sec:related-work}

\begin{table*}[!t]
\centering
\setlength\tabcolsep{0pt} 
\smallskip 

\def\arraystretch{1.2}%
\scalebox{0.95}{
\begin{tabular*}{\linewidth}{@{\extracolsep{\fill}} ccccc}
\toprule
Target & Type & Input & User Intent & \textcolor{red}{Required Chart Metadata} \\
\toprule 
    \multirow{5}{*}{Mark} 
        & \multirow{3}{*}{Select} 
            & \textbf{Mouse click} & Highlight / select mark & \multirow{2}{*}{\textcolor{red}{Marks}} \\
            & & Meta-key + mouse click & Append / un-append mark to current selection & \\
            \cline{3-5}
            & & Mouse hover & Display tooltip / details-on-demand & Marks, Legends, Axes$\rightarrow$Scales \\
      
    \cline{2-5}
  
    & \multirow{1}{*}{Filter} 
        & \textbf{Mouse click} & Filter selected mark & \textcolor{red}{Marks} \\
  
    \cline{2-5}
  
    & \multirow{1}{*}{Sort} 
        & Mouse click + mouse drag & Re-position mark to enable sorting & \textcolor{red}{Marks, Axes$\rightarrow$Scales} \\
  
  
  
  
  
    \midrule
  
    \multirow{3}{*}{Legend} 
        & \multirow{2}{*}{Select} 
            & \textbf{Mouse click} & Highlight marks by selected legend attribute & \multirow{2}{*}{\textcolor{red}{Marks, Legends}} \\
            & & Meta-key + mouse click & Append marks on legend attribute to current selection & \\
  
    \cline{2-5}
  
    & Filter & \textbf{Mouse click} & Filter marks by selected legend attribute & \textcolor{red}{Marks, Legends} \\
  
  
  
  
  
    \midrule

    \multirow{9}{*}{Background} 
        & Select & \textbf{Mouse click} & Remove current selection & \textcolor{red}{Marks} \\

    \cline{2-5}
    
    & \multirow{5}{*}{Navigate} 
        & Mouse scroll & \multirow{2}{*}{Scale view (zoom)} & \multirow{2}{*}{\textcolor{red}{Marks, Axes$\rightarrow$Scales + Views}} \\
        & & \textbf{Double click} & \\
        \cline{3-5}
        & & \textbf{Mouse click + mouse drag} & Translate view (pan) & \multirow{3}{*}{\textcolor{red}{Marks, Axes$\rightarrow$Scales + Views}}  \\
        & & \textbf{Mouse click + mouse drag} & Select area to navigate to (pan + zoom) & \\
        & & \textbf{Double click} & Reset view & \\
  
    
   
    
    \cline{2-5}
    
    & \multirow{2}{*}{Brush} 
        & \textbf{Mouse click + mouse drag} & Brush area & \multirow{2}{*}{\textcolor{red}{Marks, Axes$\rightarrow$Scales}} \\
        & & Meta-key + mouse click + drag & Append brushed area to current selection & \\
        
    \cline{2-5}
    
    
    \cline{2-5}
    
    & Annotate & \textbf{Mouse click} & Place label annotation & \textcolor{red}{Marks, Axes$\rightarrow$Scales} \\
    
    \midrule
  
    \multirow{7}{*}{Axis} 
        & \multirow{5}{*}{Navigate} & Mouse scroll & \multirow{2}{*}{Scale axis (zoom)} &  \multirow{2}{*}{\textcolor{red}{Marks, Axes$\rightarrow$Scales + Views}} \\
            & & \textbf{Double click} & \\
            \cline{3-5}
            & & \textbf{Mouse click + mouse drag} & Translate axis (pan) & \multirow{3}{*}{\textcolor{red}{Marks, Axes$\rightarrow$Scales + Views}} \\
            & & \textbf{Mouse click + mouse drag} & Select axis area to navigate to (pan + zoom) & \\
            & & \textbf{Double click} & Reset axis & \\

    \cline{2-5}

    & \multirow{2}{*}{Brush} 
        & \textbf{Mouse click + mouse drag} & Brush axis area & \multirow{2}{*}{\textcolor{red}{Marks, Axes$\rightarrow$Scales}} \\
        & & Meta-key + mouse click + drag & Append brushed axis area to current selection & \\
    
    \bottomrule
\end{tabular*}
}
\vspace{-5pt}
\caption{
Target $\times$ Types + Inputs taxonomy. We categorize user interactions by \emph{interaction target}, \emph{interaction type}, and \emph{input mechanism} for mouse and meta-key inputs. Bold text indicates ambiguous inputs relative to user intent.}
\label{table:taxonomy}
\vspace{-3mm}
\end{table*}

DIVI extends prior work on visualization tools, interaction taxonomies, and computational chart interpretation.

\subsection{Designing Interactions for Visualizations}
A variety of visualization languages and systems support custom interaction design. 
Protovis~\cite{bostock2009protovis}, D3~\cite{bostock2011d3}, and VisDock~\cite{choi2015visdock} allow developers to coordinate interactions with low-level event handlers, providing full expressive control at the cost of ease of use, especially for those without software engineering expertise.
\textcolor{red}{
While VisDock also targets SVG visualizations, DIVI handles interactions automatically rather than pre-specification (e.g., code). Further, SVG analysis enables interactions that depend on the underlying data, resulting in wider coverage of interaction intents, such as dynamic tooltips, filtering, sorting, and multi-view linking, and novel support for popular tools (ggplot2~\cite{ggplot2}). 
}

Vega-Lite~\cite{satyanarayan2016vega} provides a declarative grammar with more concise, high-level specifications. Developers can specify composable interactions as selections of points or ranges, rather than interfacing with event streams and handlers directly.
Lyra 2~\cite{zong2020lyra} is a graphical interface for custom interaction design, bridging the gap with prior work on graphical tools for static visualizations, such as Lyra~\cite{satyanarayan2014lyra}, Charticulator~\cite{ren2018charticulator}, and Data Illustrator~\cite{liu2018data}. 
Lyra 2 uses interactive \textit{demonstrations}, such as mouse click and drag events, to suggest appropriate selections and transforms (e.g., panning), which the developer can then select from a menu.
Visualization tools such as Tableau~\cite{tableau} and Plotly~\cite{plotly} provide menu or tool-palette based interactions for end users, without much customization capability.
For instance, both tools support basic panning, zooming, brushing, tooltip-on-hover, and legend filtering.
However, interactions are tightly linked to the source tool and cannot freely extend to different chart and tool compositions without manual revisions.



\subsection{Interaction Taxonomies}

Researchers have proposed various taxonomies to scaffold the interaction design space. Yi et al.~\cite{yi2007toward} categorize visualization interaction techniques into seven high-level categories: \textit{Select}, \textit{Explore}, \textit{Reconfigure}, \textit{Encode}, \textit{Abstract / Elaborate}, \textit{Filter}, and \textit{Connect}. 
Brehmer et al.~\cite{brehmer2013multi} provide a multi-level typology of abstract visualization tasks to address \textit{how} and \textit{why} a task is performed, organizing concise descriptions and comparisons of various tasks. 
In their \textit{how} typology, they provide associated interaction techniques, such as \textit{Select}, \textit{Navigate} (within Yi et al.'s \textit{Explore} category), \textit{Change} (within Yi et al.'s \textit{Reconfigure} category), and \textit{Filter}.
\textcolor{red}{Sedig et al.~\cite{sedig2013interaction} synthesize interactions into a broader theoretical framework that accounts for complex cognitive actions, such as measuring, linking, and blending.}
We use these interaction taxonomies to cover an expressive gamut of interaction \textit{types}, and to scaffold the interaction primitives in our own taxonomy (Table~\ref{table:taxonomy}).

Other systems provide their own interaction taxonomies. 
Interaction+~\cite{lu2017interaction+} provides a palette for custom selections, filters, comparisons, and annotations.
\textcolor{red}{Transmogrification~\cite{brosz2013transmogrification} supports 2D shape transform interactions and free-form prototyping. However, as with VisDock~\cite{choi2015visdock}, these tools do not access or infer underlying data required for exploratory interaction (e.g., tooltips, linking, data filtering / sorting, and axis-based zooming)}.
InChorus~\cite{srinivasan2020inchorus} proposes a multimodal interaction taxonomy for tablet-based visualizations.
While multimodal input can resolve gestural ambiguities~\cite{wobbrock2009user}, most InChorus interactions are realized as custom combinations of interaction primitives. 
For example, one must sort a bar chart using specific voice commands or swiping along an axis, rather than dragging the smallest / largest bars as a demonstration.
Here, we apply DIVI to traditional (mouse / keyboard / touchpad) modalities to test its usability with a small set of familiar interaction primitives.
Using multimodal input to resolve ambiguities in DIVI remains as interesting future work.

\subsection{Programming by Demonstration (PBD)}
Programming by Demonstration (PBD) tools construct programs from user input demonstrations~\cite{mcdaniel1999getting} and may improve a design workflow, as in Lyra 2~\cite{zong2020lyra}.
Snap-Together Visualization~\cite{north2000snap} applies PBD to multiple coordinated views, allowing end users to dynamically combine visualizations and configure interactions such as details-on-demand and linked brushing. 
Falx~\cite{wang2021falx} and Saket et al.'s Visualization by Demonstration~\cite{saket2016visualization, saket2019investigating, saket2019liger} explore how interactive demonstrations (e.g., re-positioning points in a scatter plot) can be used to infer and refine global transforms to the visual design (e.g., bin points in close proximity together), removing the need to provide manual or textual design specifications.
These systems focus on chart specification, whereas DIVI concerns subsequent selection and view transformation.
We take inspiration from this thread of research, exploring how specifications of interactions can be automatically inferred at runtime.

\subsection{Deconstructing Visualizations}
Deconstructing visualizations to extract meaningful chart information---such as marks, data attributes, and visual encoding channels---can support chart analysis, classification, and design re-targeting.
ReVision~\cite{savva2011revision}, Graphical Overlays~\cite{kong2012graphical}, REV (Reverse-Engineering Visualizations)~\cite{poco2017reverse, poco2017extracting}, ChartSense~\cite{jung2017chartsense}, and approaches surveyed with Chart Mining~\cite{davila2020chart} use computer vision techniques to automatically extract marks and infer visual encoding channels and data attributes from bitmap images.
\textcolor{red}{However, these techniques suffer from a variety of issues that affect inference precision, such as pixel resolution, shape ambiguity (e.g., blurry edges), and mark occlusion~\cite{masson2023chartdetective}.}
While our work targets structured SVG images, these techniques might enable DIVI to support a broader set of visualization media types in the future.

DIVI's SVG analysis follows prior work on deconstructing D3 visualizations~\cite{harper2014deconstructing, harper2017converting}, as well as Reviz~\cite{reviz}, which produces partial visualization programs for re-targeting with Observerable Plot~\cite{observable}.
These tools exploit D3's~\cite{bostock2011d3} data-binding protocol to extract marks and data for restyling, 
\textcolor{red}{
but do not infer axes and legends or generalize to charts produced by tools other than D3.
Liu et. al~\cite{liu2023spatial} contribute a spatial-constraint model to enable interaction with static visualizations, although its deconstruction output is limited to evenly spaced axes (prohibiting use of log scales) and marks, without support for linked interaction and other non-spatial exploratory interactions (e.g., tooltips and selection / brushing).
ChartDetective~\cite{masson2023chartdetective} provides semi-automated data extraction methods for vector charts in PDFs, but lacks support for chart metadata (such as axes) needed for interaction and linking.
In addition, semi-automated approaches requiring extensive user input are not as scalable. Instead, users can correct DIVI's automatic methods by exposing certain metadata directly just in cases where inference fails.
}

\section{Formative Discussions with \textcolor{red}{Data Analysts}}
\label{eval:interviews}

To inform DIVI, we conducted discussions with 6 experts in statistics and data analysis to motivate the need for interaction decoupling, allowing interactions to extend to more common visualization tools in data analysis that output static charts.
We interviewed PhD students (2), scientific researchers (3), and an industrial data scientist (1), each for 30 minutes.
\textcolor{red}{Participants primarily worked in analysis environments (R or Jupyter) and so gravitated to the visualization tools common to these environments (e.g., ggplot2~\cite{ggplot2}, Matplotlib~\cite{matplotlib}), many of which do not provide interaction support.
Interviews were semi-structured. We first collected qualitative feedback on (1) analysis tasks for which visualization is helpful; (2) commonly used data analysis and visualization tools; (3) types of interactions most often used or desired; and (4) issues frequently encountered when using or designing interactions.
We then demonstrated dynamic interaction examples in DIVI to prompt open-ended discussion and solicit feedback.}

Participants reported regularly using the following tools to create visualizations: Matplotlib / Seaborn (4), Vega-Lite / Altair (1), D3 (1), and R / ggplot2 (3).
Most participants (5) rarely use interactive charts, using tools like Plotly~\cite{plotly} when basic interactions are needed.
\textcolor{red}{No participants reported designing interactions themselves.}
The reasons for low adoption of interactive plots include the lack of interactive support in tools such as Matplotlib~\cite{matplotlib} and ggplot2~\cite{ggplot2}, technical barriers or lack of modularity associated with moving from interaction specification (e.g., Vega-Lite~\cite{satyanarayan2016vega}) to procedural code (e.g., D3~\cite{bostock2011d3}), and some concerns about the usefulness of interaction.
Most participants use manual chart re-plotting as a proxy for interaction; e.g., modifying axis extents in the chart specification in lieu of zoom or pan interactions, changing grouping fields, or faceting.
Consequently, most participants (5) listed faceted views with linked selections as a highly desired feature, in order to quickly sift through different subsets of the data that are often impractical to visualize in one view.
Overall, participants found interaction underutilized in practice since it is not typically possible to map interactive selections back to the original data for further analysis.

Participants were enthusiastic about potential applications of DIVI, in particular: recover data from charts (i.e., download inferred data), maintain an interaction history, bookmark view snapshots, or map back to rows of the original table.
\textcolor{red}{
In addition, participants noted uses for presentation-level graphics, re-targeting across tools, and interaction with vectorized PDFs.
For example, extracted chart metadata might be used to highlight interesting data during a presentation, shift between different visualization tools, or interact with PDF charts and text while reading scientific papers.
Two participants stated that automatically applying DIVI to rendered output in Jupyter notebooks would make it especially attractive for their exploratory analysis workflows.
}

\textcolor{red}{
These formative discussions motivated several design goals (\S\ref{sec:design-goals}), in particular the need for dynamic interaction rather than manual interaction specification (DG3) and interaction decoupling (DG2) to support popular analysis environments (e.g., R and Jupyter) where exploratory interaction remains largely unsupported for novice visualization users. 
In addition, we used DIVI's decoupling to enable linked interactions, a highly desired feature (5/6 participants) for exploring multi-dimensional data typical of real-world analysis settings.
}

\textcolor{red}{\section{Design Goals}
\label{sec:design-goals}

\begin{figure*}[!t]
    \centering
    \resizebox{0.96\linewidth}{!}{\includegraphics{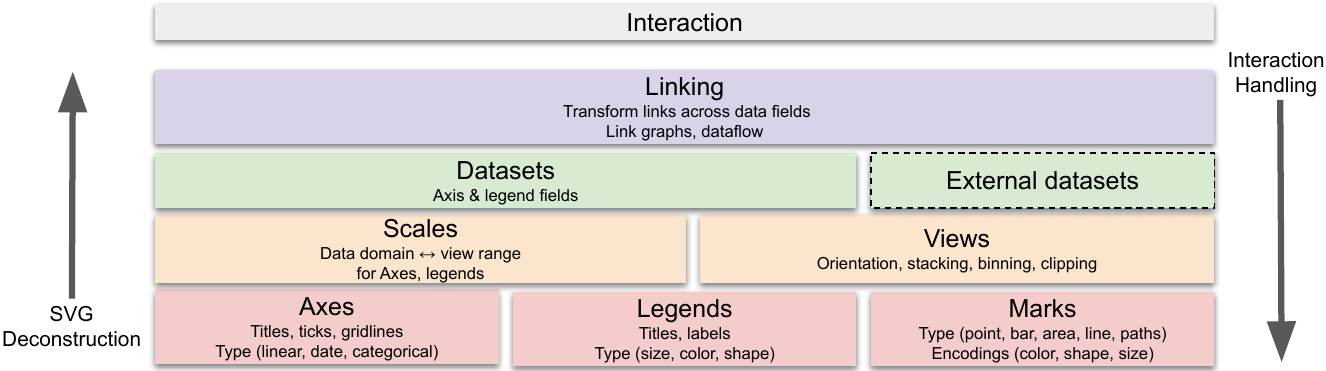}}
    \vspace{-5pt}
    \caption{DIVI's inference model and outputs, enabling interaction decoupling from chart specification. SVG charts are deconstructed to identify core chart structures: axes, legends, and marks. Scales, view information, data fields, and linkings are then computed. Downstream interaction logic propagates via link graphs and materializes back to SVG elements.}
    \label{fig:svgstack}
    \vspace{-3mm}
\end{figure*}

DIVI is intended to automatically support sufficient exploratory interactions without prior specification.
Based on our analyst interviews, interaction taxonomies~\cite{yi2007toward,brehmer2013multi}, and design tools~\cite{zong2020lyra,choi2015visdock,lu2017interaction+,srinivasan2020inchorus,wobbrock2009user}, we identified four design goals to motivate DIVI's approach:

\textbf{DG1: Support an expressive, relevant set of interactions.}
Visualization users require a useful set of interactions to explore and analyze data depending on the task at hand, a sentiment widely shared in our discussions with data analysts (\S\ref{eval:interviews}).
DIVI uses prior interaction taxonomies~\cite{yi2007toward} to ensure coverage over an expressive and applicable set of exploratory interactions for a variety of chart types and user goals (select / brush, filter, sort, annotate, and navigate) (\S\ref{interaction}).

\textbf{DG2: Decouple interaction from source tools.}
Interaction is often tightly coupled with the design process of a visualization. 
By decoupling interaction from this process to standardize chart metadata (e.g., axes, scales, etc.), analysts and designers can focus on visual encoding first and select a range of possible interactions later.
Additionally, this decoupling enables interactions for popular tools without native interaction support (e.g., ggplot2~\cite{ggplot2} and Matplotlib~\cite{matplotlib}).
Linked interactions, including across charts produced by different visualization tools, can also be supported directly.

\textbf{DG3: Enable dynamic interaction rather than prior interaction specification.}
DIVI allows users to directly interact with charts in lieu of writing interaction handling code, an often tedious and/or technically demanding process.
DIVI uses mediation techniques~\cite{dey2005designing} to clarify ambiguous interaction intents (\S\ref{interaction:handling}).
We contribute a taxonomy (\S\ref{interaction}) to delineate the interaction space across target chart elements and input events, as well as the required chart metadata, which plotting tools can expose directly in the future to support dynamic interaction.

\textbf{DG4: Leverage SVG deconstruction to automatically infer chart semantics.}
DIVI uses sufficient chart metadata to support dynamic interaction (DG3). 
While perhaps theoretically possible, updating all popular visualization tools to expose metadata for interaction decoupling (DG2) is infeasible in the short term.
Rather, DIVI automatically infers chart semantics (axes, legends, etc.) from SVG, which is sufficiently rich to enable this analysis and modify for interactive updates.
In addition, given chart metadata and access to underlying data (either directly, or via extraction from non-aggregated plots (\S\ref{deconstruction:datasets})), DIVI can infer requisite linked interactions across views.

}
\section{\textcolor{red}{Interaction Taxonomy}}
\label{interaction}


To delineate standard user interactions for dynamic orchestration, we formulated the interaction taxonomy of Table~\ref{table:taxonomy}.
While researchers have surveyed \textit{types} of interactions such as filtering and zooming~\cite{brehmer2013multi,yi2007toward}, these treatments stop short of meticulously mapping the various input events by which those interactions are realized. 
This lower level of detail is necessary for recognizing common user inputs and coordinating resulting view transforms to support dynamic interaction. 
Critically, our mapping of input events also outlines the space of overlapping interactions (e.g., drag to pan vs. brush) that DIVI uses for disambiguating user intents (DG3).
Accordingly, our taxonomy focuses on the following questions:
\textit{What is the space of possible interaction targets (i.e., visual elements) in a visualization?}
\textit{Given a target visual element, what are the possible interaction types?}
\textit{And given an interaction type, what are the possible input events?}
\textcolor{red}{\textit{What chart information is needed for a given interaction?}}
In the following paragraphs, we outline each of these dimensions in the taxonomy and corresponding chart metadata.

\textbf{Interaction Type}.
We base our interaction types on Yi et. al's~\cite{yi2007toward} seven general categories \textcolor{red}{to ensure coverage over an expressive, relevant set of exploratory interaction techniques (DG1)}, noting the correspondence in parentheses.
\textbf{Select} (\textit{Select}) refers to the inspection of individual data items, such as graphical marks on a chart, and includes details-on-demand (e.g., tooltip) and highlighting.
\textbf{Filter} (\textit{Filter}) interactions involve removing selected (exclusive) or unselected (inclusive) marks. 
\textbf{Sort} (\textit{Reconfigure}) interactions modify the data ordering based on a subset of the data attributes (e.g., order bar chart by ascending $y$ values).
\textbf{Annotation} (\textit{Select}) interactions enable end users to label items of interest (e.g., annotation with text labels and arrows).
\textbf{Navigate} (\textit{Abstract / Elaborate}) interactions scale (zoom) or translate (pan) the view to focus on a different subset of the data.
\textbf{Brush} (\textit{Select}) refers to selections of a subset of items, such as area-based selections in a single view or linked selections across multiple views.

\textbf{Interaction Target \& Input Event}.
We identified graphical marks, chart background, legends, and axes as interaction targets.
DIVI supports mouse / touchpad and keyboard input.
We support mouse click (inc. right-click), double click, hover, drag, scroll, and keyboard meta-keys (shift, control, command, etc.).

\textcolor{red}{\textbf{Required Chart Metadata}.
We identified axes, legends, and marks as chart components required to materialize interaction. DIVI infers these components (\S\ref{deconstruction}) to reconstruct further information (e.g., datasets and multi-view links) and drive interaction logic. We outline which metadata are needed for each single-view interaction target and type: marks are required for all interactions, legends for all legend and tooltip interactions, and axes for all interactions except mark selection and filter. Arrows ($\rightarrow$) indicate required metadata that can be derived from marks, axes, or legends. For example, after deconstructing axes, DIVI derives scale and view metadata for navigation (i.e., rescaling marks / axes and clipping). Each interaction may be linked across views, requiring all three components to reconstruct datasets and infer linkage types 
(i.e., Marks, Axes + Legends$\rightarrow$Datasets$\rightarrow$Linking) (\S\ref{deconstruction},~\ref{linking}).
}

\section{SVG Deconstruction}
\label{deconstruction}

\begin{figure*}[t!]
    \centering
    \resizebox{0.96\linewidth}{!}{\includegraphics{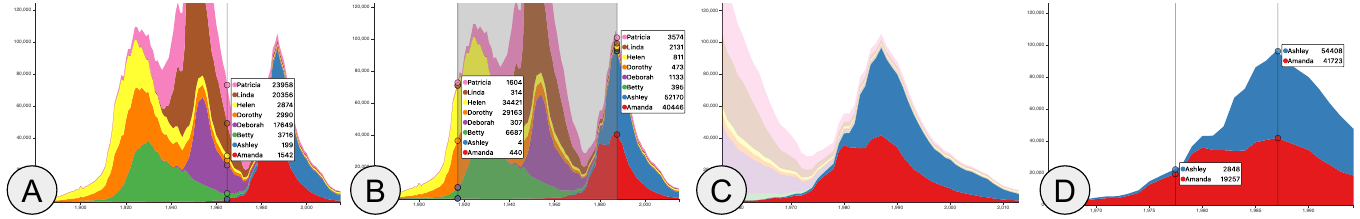}}
    \vspace{-5pt}
    \caption{DIVI with a D3 stacked area chart. Left to right: (A) hover tooltips show data series values, (B) brushing displays tooltips for the start and end points, (C) areas selected via clicking or brushing, and (D) areas filtered then brushed selectively.}
    \label{fig:stackedarea}
    \vspace{-3mm}
\end{figure*}

\begin{figure}[!b]
    \centering
    \resizebox{1\linewidth}{!}{\includegraphics{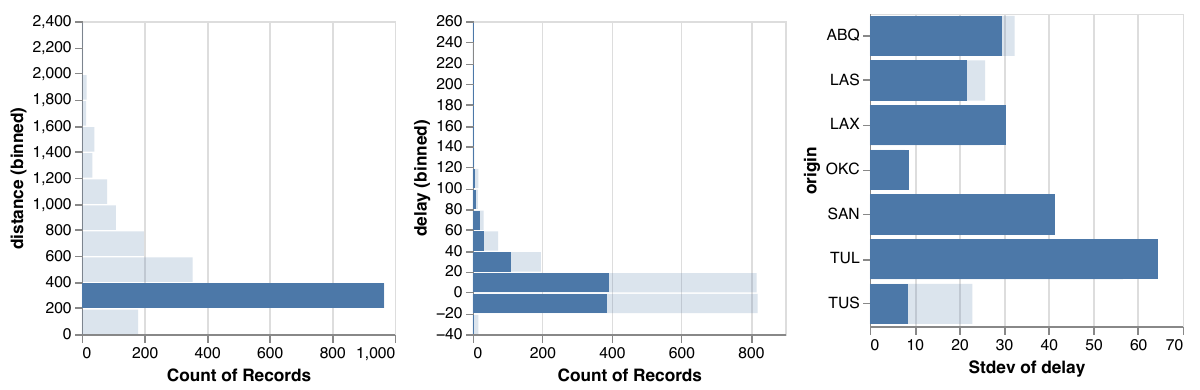}}
    \vspace{-15pt}
    \caption{Linked crossfiltering of Vega-Lite histograms of flight data, inferred from SVG output and a single provided data table. The table consists of the number of records on binned \texttt{distance} (left), number of records on binned \texttt{delay} (middle), and \texttt{stddev} of \texttt{delay} by \texttt{origin} filtered to 6 airports. Clicking on the $[200, 400]$ distance bin (left) selects corresponding backing flight data and re-aggregates dependent views (middle, right). In this example, the \texttt{stdev} of \texttt{delay} for most of the selected airports appears highest for low \texttt{distance}.}
    \label{fig:crossfilter}
    \vspace{-1mm}
\end{figure}

To add dynamic interaction to static charts, we first interpret chart structure and semantics.
DIVI performs SVG deconstruction to identify semantic components of a visualization and uses this information to automatically (1) infer data, (2) link data across views, and (3) coordinate subsequent user interaction with these components.
We discuss data inference in this section, linking in \S\ref{linking}, and interaction handling in \S\ref{interaction:handling}.

DIVI deconstructs SVG charts in three sequential phases (Fig.~\ref{fig:svgstack}): 
    \begin{itemize}
        \item Mark-level analysis to infer axes, legends, marks, and titles
        \vspace{-4pt}
        \item View-level analysis to reconstruct scales and view encodings
        \vspace{-4pt}
        \item Data-level analysis to determine relational table fields
    \end{itemize}
DIVI is a JavaScript library, which visualization developers can use to post-process one or more static SVG charts. 
Critically, no manual interaction code is needed. 
Deconstruction and interaction handlers are instantiated and initialized with the single function call \sloppy\texttt{DIVI.hydrate([node1, node2, ...], data)}, which takes one or more DOM (Document Object Model) nodes or corresponding CSS selectors and an optional data source (for linking in \S\ref{linking}) as input.
In the following subsections, we detail each deconstruction phase. 

\subsection{Mark Analysis}
\subsubsection{Inspection}
\label{deconstruction:inspection}
DIVI's inspection phase parses the SVG DOM tree to identify DOM elements and features needed for analysis in the next sections.
For each input chart DOM element, DIVI performs a pre-order traversal of the DOM subtree to extract individual marks. 
DIVI makes no assumptions about the layout or structure of the DOM tree to decouple from source tools. 
As each child is accessed, DIVI first inspects the element type. 
SVG elements (\texttt{svg}) are maintained for screen / SVG coordinate conversions, and group elements (\texttt{g}) with transforms (e.g., {\texttt{\textless g transform="translate(50,10)"\textgreater}}) are extracted to maintain global and local translation, scale, and rotation matrices. These matrices are bound to mark elements for efficient view manipulations during interactions such as zooming (\S\ref{interaction:handling}).

DIVI recurses over child elements.
We collect base SVG and text marks (\texttt{circle}, \texttt{ellipse}, \texttt{line}, \texttt{polygon}, \texttt{polyline}, \texttt{rect}, \texttt{path}, \texttt{use}, and \texttt{text}) 
and store the element type for subsequent axis, legend, and mark analysis. 
We categorize \texttt{path} elements as lines, circles / ellipses, rectangles, or generic polygons by inspecting path commands.
For instance, the \texttt{d} attribute for polygons will end with a \texttt{Z} (close path) command or \texttt{M} (move to) command that ends at the starting point.
Polygons with a bounding box dimension that matches an axis are categorized as area marks (as in the stacked area chart in Fig.~\ref{fig:stackedarea}).

\subsubsection{Axes}
\label{deconstruction:axes}
DIVI assumes each axis tick or gridline grouping shares the same width, height, color, and mark type (\S\ref{deconstruction:inspection}) \textcolor{red}{and uses positional relations to identify axis candidates (Appendix~\ref{appendix:grouping},~\ref{appendix:axes}).}
Without this assumption, marks aligned with candidate ticks or gridlines 
may be incorrectly labeled as orphan ticks (i.e., ticks without a corresponding text label).
\textcolor{red}{
We have not yet encountered tools outputting axes without shared styling given the perceptual illegibility that would occur.
}

\subsubsection{Legends \& Titles}
\label{deconstruction:legends}
Text marks that are not grouped with axis tick marks are analyzed to determine if they are axis titles or legend keys again using positional relations and styling \textcolor{red}{(Appendix~\ref{appendix:legends})}.
DIVI supports color, size, and shape legends--and their combinations.

\subsubsection{Marks}
\label{deconstruction:marks}
Remaining marks outside of any axis extent are identified as \texttt{viewport} instances.
For example, ggplot2~\cite{ggplot2} and Matplotlib~\cite{matplotlib} may include bounding boxes or lines as axis edges (Fig.~\ref{fig:teaser}).
Any text or mark that is not identified within an axis, legend, title, or as a viewport instance is stored as a mark.
Marks are specifically used for subsequent view and data analysis, as well as interaction targets (Table~\ref{table:taxonomy}).

\subsection{View Analysis}
\subsubsection{Scales}
\label{deconstruction:scales}
Scales are critically needed to map from the data space to screen space (or vice versa), driving data inference and view transformations such as axis rescaling during navigation.
DIVI currently supports continuous linear and logarithmic (log) scales, categorical / ordinal scales, and date scales for deconstructed axes \textcolor{red}{using corresponding tick labels (Appendix~\ref{appendix:scales})}.
Deconstructed color or shape legends are stored as key-value mappings from the respective mark encoding to the legend label.
We construct linear scales for size legends, supporting sizes encoded as changes of mark width, height, or area.

\subsubsection{Views}
\label{deconstruction:views}
DIVI infers global chart arrangements from deconstructed axes and marks: orientation, stacking, and binning \textcolor{red}{(Appendix~\ref{appendix:views})}.
Axis extents for all charts are used to instrument SVG clipping transforms for subsequent interaction (e.g., navigation).

\subsection{Data Analysis}
\label{deconstruction:datasets}
DIVI inverts reconstructed scales from axes and legends to infer datasets \textcolor{red}{(Appendix~\ref{appendix:data-analysis})}.
These datasets drive DIVI's core interaction logic: datasets are used for link inference (\S\ref{linking}) and interaction handling (\S\ref{interaction:handling}) to propagate selections and transforms at the data-level.
While users may have access to the data used to generate the SVG visualization, data inference is necessary to link marks to their original data attributes.
Without inference, users must either manipulate the marks themselves with custom handlers (e.g., D3~\cite{bostock2011d3, choi2015visdock}) or use specific tools (e.g.,~\cite{satyanarayan2016vega}), violating DIVI's core goal of interaction decoupling.
\textcolor{red}{DIVI's final deconstruction output consists of the following metadata: 
\begin{itemize}
    \item \sloppy\texttt{svgContainer}, \texttt{chartTitle}, \texttt{marks}, \texttt{dataTable}
    \item \sloppy\texttt{axes}: \texttt{tickLabels}, \texttt{tickMarks}, \texttt{scale}, \texttt{title}, \texttt{orientation}, \texttt{bins}
    \item \sloppy\texttt{legends:} \texttt{type}, \texttt{scale}, \texttt{title}, \texttt{legendLabels}, \texttt{legendMarks}
\end{itemize}
}
\textcolor{red}{In the future, charting tools could expose this information to support dynamic interaction directly,  without requiring inference.}

\section{Multi-view Linking}
\label{linking}

\begin{figure*}[!t]
    \centering
    \resizebox{0.95\linewidth}{!}{\includegraphics{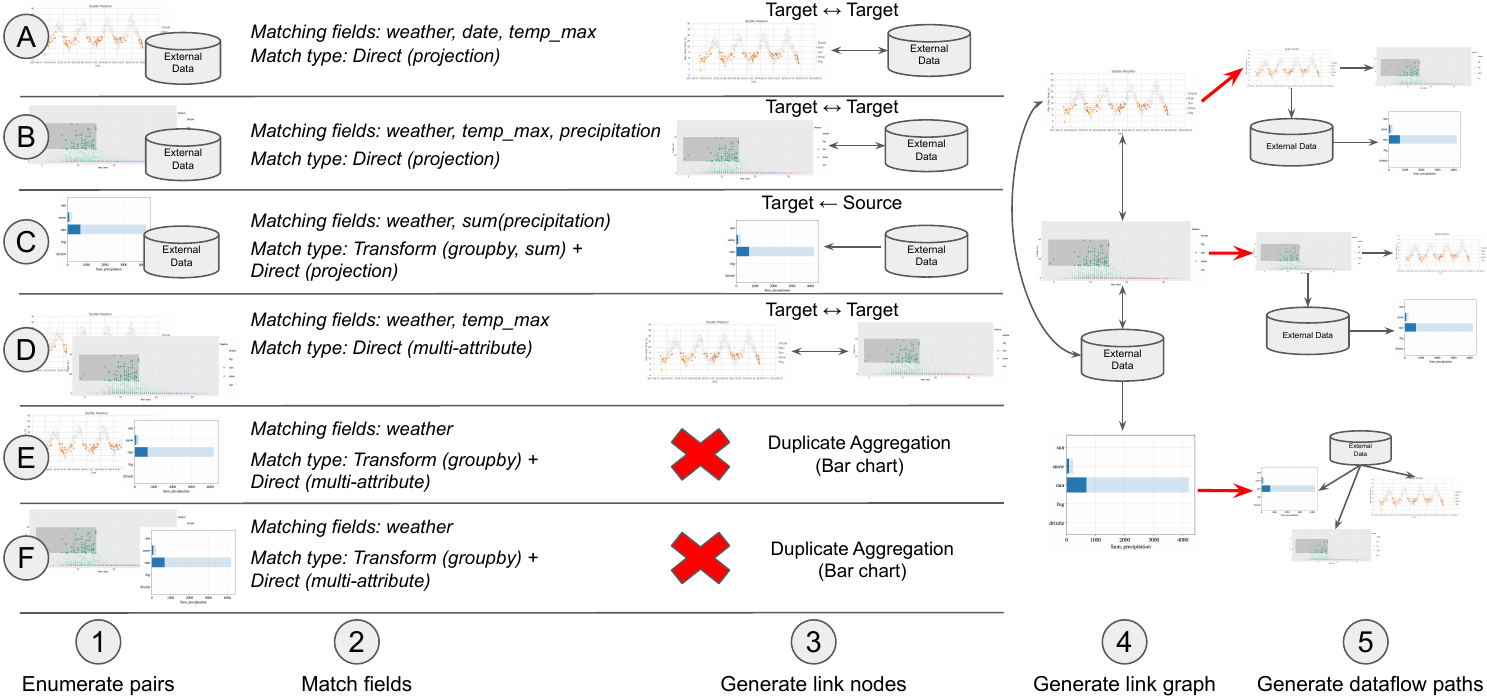}}
    \vspace{-5pt}
    \caption{Link search procedure. DIVI (1) enumerates pairs of visualizations, first checking against external data tables if they are provided. (2) Matching fields and link types are identified via \textit{relationship linking} and \textit{transform linking} subroutines. (3) \texttt{Target} and \texttt{Source} link nodes are generated based on match types. In this example, the scatter plots maintain each other and external data as \texttt{target} nodes (i.e., they update local tables before updating \texttt{target} tables), while the bar chart maintains the external data as a \texttt{source}, meaning that any interactions (e.g., selection) first select from \texttt{source} tables before updating their own, as in the case of re-aggregation. Since the bar chart matches on all fields with a transformation applied to the external data (\texttt{groupby} + \texttt{sum} aggregation), it ignores adding the scatter plots as \texttt{source} nodes since they only match one field. (4) DIVI generates a single link graph based on discovered link node relations. (5) DIVI specifies a dataflow path for each chart when users interact with it by first recurisvely updating any \texttt{source} nodes, then propagating to all \texttt{target} nodes. }
    \label{fig:dataflow}
    \vspace{-3mm}
\end{figure*}

After performing SVG analysis to infer individual chart semantics, DIVI attempts to determine possible multi-view coordination schemes if multiple charts were provided as input. 
DIVI infers links between views in a two-step process: (1) link inference between visualization pairs, and (2) generation of link graphs to specify dataflow logic. 

\subsection{Inference}
DIVI enumerates pairs of visualizations $v_1$ and $v_2$ with corresponding data tables (datasets) $D_1$ and $D_2$ to check for potential links.
If one chart visualizes all data points in a table directly, a full dataset can be recovered from the SVG alone.
For other cases, an external dataset (URL or file) may be included as an additional data table.
This provision is necessary when inference is impossible without access to the original data (e.g., cross-filtering between aggregated views only, as in Fig.~\ref{fig:crossfilter}).

\textbf{Linking Model}.
For candidate data tables $D_1$ and $D_2$ with respective fields $A = [a_1, a_2, ..., a_n]$ and $B = [b_1, b_2, ..., b_m]$, we specify a \textit{relationship linking} $RL(D_1, D_2)$ as:
\begin{align*}
    a_i \in A, b_j \in B : D_1[a_i] \subseteq D_2[b_j]
\end{align*}

This formalism covers many common types of \textit{direct} linkings: those without transformations applied to data fields.
    \begin{enumerate}
        \item Projected views (i.e., $D_1$ uses a subset of $D_2$ columns):
            \begin{align*}
                |A| &< |B| \\
                \forall a_i \in A, \exists b_j &\in B : D_1[a_i] = D_2[b_j]
            \end{align*}
        \item Filtered views (i.e., $D_1$ uses a subset of $D_2$ rows):
            \begin{align*}
                |A| &= |B| \\
                \forall a_i \in A, \exists b_j &\in B : D_1[a_i] \subset D_2[b_j]
            \end{align*}
        \item Filtered and projected views (i.e., $D_1$ uses a subset of $D_2$ rows and columns):
            \begin{align*}
                |A| &< |B| \\
                \forall a_i \in A, \exists b_j &\in B : D_1[a_i] \subset D_2[b_j]
            \end{align*}
        \item Multi-attribute views (i.e., one or more fields in either view link together, such as US state names or dates (Fig.~\ref{fig:teaser})):
            \begin{align*}
                \exists a_i \in A, \exists b_j &\in B : D_1[a_i] = D_2[b_j]
            \end{align*}
    \end{enumerate}

For candidate transformation mappings $T = \{t_1, t_2, ..., t_k\}$ and a transform sequence $S = [s_1, s_2, ..., s_r]$ where $s_i \in T$, we specify a \textit{transform linking} $TL(D_1, D_2)$ as the following:
\begin{align*}
    D' &= [s_1(d_1), s_2(d_2), ..., s_r(d_r)] 
    \\ \text{ where } [d_1, &d_2, ..., d_r] \subseteq D_1 
    : RL(D', D_2) \neq \emptyset \\
\end{align*}

\vspace{-10pt}
\noindent
Concretely, this definition specifies some set of transformed fields from $D_1$ to yield a new table $D'$ that \textit{directly} links to $D_2$.
We consider grouping, aggregation (i.e., \texttt{min}, \texttt{max}, \texttt{mean}, \texttt{sum}, \texttt{count}, \texttt{stdev}, \texttt{median}), and derivation transforms for date formatting (e.g., removing the year from the date for grouping) and numeric binning.

\begin{figure*}[!t]
    \centering
    \resizebox{\linewidth}{!}{\includegraphics{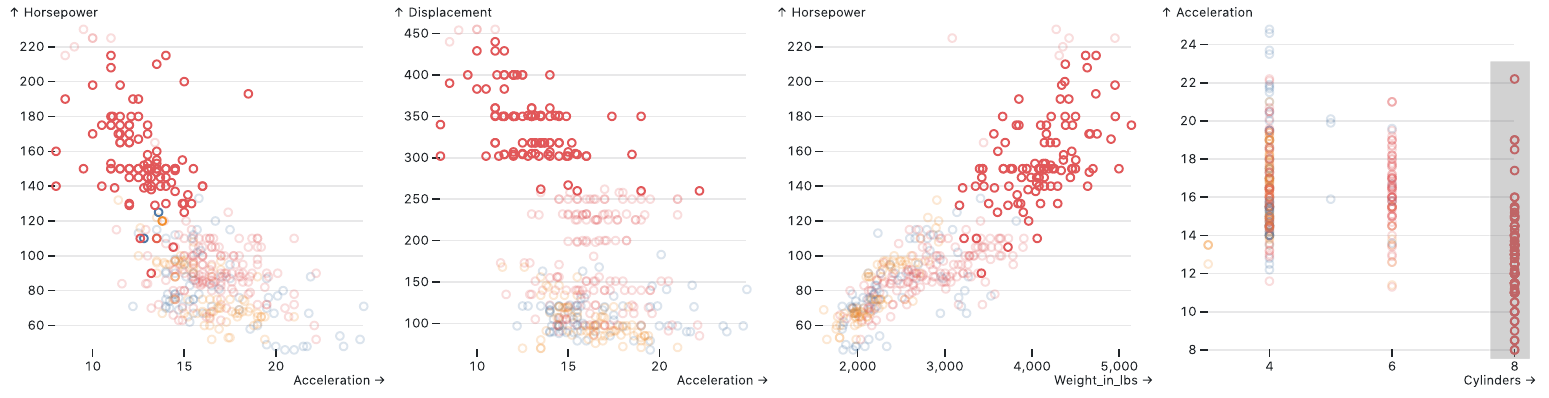}}
    \vspace{-15pt}
    \caption{Linked brushing with four static Observable Plot~\cite{observable} scatter plots (external dataset used for linking). Brushing data by cylinder groups highlights useful relations: More cylinders associates with higher weight, horsepower, and displacement, but lower acceleration.}
    \label{fig:vegalite-obplot}
    \vspace{-3mm}
\end{figure*}

\textbf{Link Search}.
Using the definitions for \textit{relationship} and \textit{transform} linkings, we operationalize \textit{linking} $L(D_n, D_m)$ as an iterative search process over candidate transform sequences.

DIVI begins the \textit{linking} procedure by enumerating pairs of visualizations and their datasets, using any externally provided data first (Fig.~\ref{fig:dataflow}(1)).
For each pair of visualizations $(v_n, v_m)$ with datasets $(D_n, D_m)$,
DIVI first checks a \textit{relationship linking} $RL(D_n, D_m)$ on the original, non-transformed tables, performing semi-joins on candidate columns. 
We allow an $\epsilon$-error of $1\%$ of the data range (i.e., $\epsilon=0.01 * (\texttt{max}(D_m[b_j]) - \texttt{min}(D_m[b_j]))$) that may result from data inference (\S\ref{deconstruction:datasets}).
If \textbf{every} $D_m[b_j]$ value is matched, the candidate column pair ($a_i, b_j$) is added as an attribute projection link and removed from the search set, along with non-matching rows in $D_n$.
If a \textit{relationship linking} is found on the non-transformed tables, the linking process terminates.
An example of this process can be seen in Fig.~\ref{fig:dataflow}(2A).
The first pair of visualizations is the Excel scatter plot and external weather data (from Fig.~\ref{fig:teaser}). 
Since the scatter plot \textit{projects} the \texttt{weather} (legend), \texttt{date} ($x$-axis), and \texttt{temp\_max} ($y$-axis) fields for all rows in the original data, the $RL(D_n, D_m)$ process matches the arrays for these 3 fields, terminating since all fields of the scatter plot have been exhausted. 
As at least one field matches, the linking process ends, outputting the field name pair and external dataset for each matched array.

If a \textit{relationship linking} is not found initially, DIVI enumerates candidate transforms (\texttt{groupby}, \texttt{aggregate}, and \texttt{derive} for dates and numeric binning) for each data field (i.e., \textit{transform linking}), applies the candidate transform as a query to the backing data table, and recursively checks for a \textit{relationship linking}.
An example of this can be seen in Fig.~\ref{fig:dataflow}(2C), consisting of the Matplotlib bar chart and external dataset. 
DIVI's recursive enumeration eventually applies \texttt{groupby} on \texttt{weather} and a \texttt{sum} aggregation on \texttt{temp\_max} in the original dataset. 
Because this new table has a \textit{relationship linking} with the bar chart (i.e., the two fields in both datasets have equal arrays), the process terminates, outputting the matched fields and data transformations that were used in the enumeration process. See Appendix~\ref{linking-algorithm} for \textit{linking} pseudocode.

\subsection{Link Dependency Graphs}
\label{linking:graphs}
DIVI generates dataflow graphs from inferred links to propagate interaction logic (e.g., selections) between linked views (Fig.~\ref{fig:dataflow}(3-5)).
This graph is represented with \texttt{target} and \texttt{source} nodes between pairs of view states / data tables.
A \textit{relationship linking} without transforms (i.e., \textit{direct} linking) between visualization pairs adds each view's state as a \texttt{target} of the other.
As an example, Fig.~\ref{fig:dataflow}(2A, B, D) each have a direct relationship (i.e., there is no transformation applied).
As a result, each pair adds the other as a \texttt{target} node.
This decision reflects that any update to one of these visualizations should update its data table before propagating to other linked (\texttt{target}) tables.
For a \textit{transform linking} between $(v_1, v_2)$, the linking direction specifies a \texttt{source}-\texttt{target} relationship in the graph.
For example, brushing an aggregated view (Fig.~\ref{fig:crossfilter}) should first select the backing data in the \texttt{source} view, re-aggregate the data, and render the new data to any \texttt{target} nodes.
As show in Fig.~\ref{fig:dataflow}(3C), the aggregated bar chart specifies the external data table as a \texttt{source} node: any selections (e.g., brushing) will first update the external data table and re-aggregate the selected data as overlaying bars.
If DIVI encounters duplicate aggregation (e.g., the bar chart also links to the scatter plots on the \texttt{weather} field), it greedily selects the aggregation scheme matching the most columns.

After specifying \texttt{source}-\texttt{target} relationships between views, dataflow paths are generated for each view by recursively selecting all \texttt{source} nodes to identify a \texttt{root} node, specifying the path of data selection and transforms (e.g., selection, brushing, zooming) when interacting with a given view.
An initial interaction updates the \texttt{root} data table.
For each \texttt{target}, the updated \texttt{source} table is transformed with inferred queries if a \textit{transform linking} exists.
The resulting table rows (and field names) are then recursively mapped to each \texttt{target}'s table rows using the row indices that matched during the \textit{relationship linking} subroutine.
An example of this process can be seen in Fig.~\ref{fig:dataflow}(5): interacting with the bar chart (e.g., brush selection) first updates its sources (the external data); the updated external data is re-aggregated when it is propagated to the bar chart, but directly linked with corresponding records in the scatter plots (since no transformation exists between them).
Interacting with one of the scatter plots first updates itself, then updates the other scatter plot and external data, and finally re-aggregates the external data for the bar chart.

\subsection{Event Handling}
\label{interaction:handling}

DIVI binds event handlers to view targets (Table~\ref{table:taxonomy}) to coordinate user input, which can be interactively disambiguated for conflicting inputs (\S\ref{disambiguation}).
Interactions are transformed into either \textit{selection} or \textit{transform} predicates: (\texttt{field}, \texttt{op}, \texttt{value}).
These predicates specify the table \texttt{field} to match on; operation type for field comparison, ordering, or view transformation (\texttt{op}); and \texttt{value} to compare, order, or transform by.
Predicates propagate across link graphs, update local data tables, and materialize as position or style transforms to core chart structures -- axes, legends, and marks. 
Here we describe predicates and SVG materialization for each interaction type.

\textbf{Select \& Brush}.
Hovering over a mark displays a tooltip with the mark's inferred data attributes.
When a mark is clicked, DIVI generates a \textit{selection} predicate that indexes into the table row maintained by the mark (Appendix~\ref{appendix:data-analysis}): (\texttt{field}: \texttt{INDEX\_FIELD}, \texttt{op}: ``='', \texttt{value}: \texttt{INDEX\_VALUE}).
For legend marks, DIVI generates the predicate (\texttt{field}: \texttt{LEGEND\_FIELD}, \texttt{op}: ``='', \texttt{value}: \texttt{LEGEND\_VALUE}).

A \textit{selection} predicate is then transformed into a select query on the data table, recursively propagating to any linked views (\S\ref{linking:graphs}).
Queried selections are maintained as a separate table copy; additional selections operate on this table, enabling composable updates (e.g., select from a current legend selection).
We use composable selections to enable brushing. Brush selections are converted to ``<='' and ``>='' operators corresponding to the bounding box extents (mapped to the data space) and applied as sequential queries prior to link propagation.
If a meta-key (Cmd, Ctrl, Shift) is also pressed, the queried selection is instead appended to the current table copy, de-duplicated by row index.
For size legends, meta-key interaction identifies \texttt{min} and \texttt{max} legend labels and generates two composable selections (``<='' and ``>='') similar to brushing.
After selections finish propagating, DIVI iterates over updated data tables, reducing the opacity of unselected marks (Figure~\ref{fig:vegalite-obplot}).

\begin{figure*}[!b]
    \centering
    \resizebox{\linewidth}{!}{\includegraphics{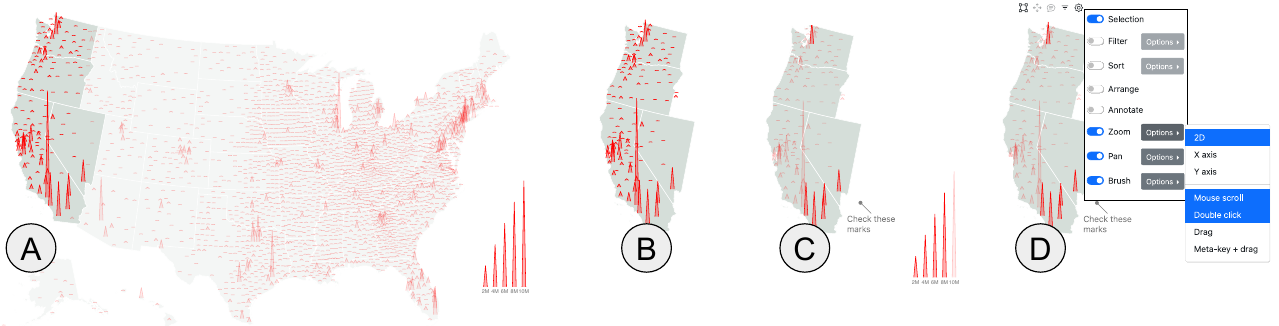}}
    \vspace{-15pt}
    \caption{DIVI with a D3 map of US county populations. (A) Specific marks (e.g., states and population markers) can be brushed and (B) inclusively filtered. (C) Clicking items in the size legend creates range-based selections, an annotation documents an observation to share. (D) Users can invoke a global menu to further customize or limit the provided interactions.}
    \label{fig:map}
    \vspace{-1mm}
\end{figure*}

\textbf{Filter}.
Selected marks (either view or legend marks) can be filtered inclusively or exclusively using the filter icon (Figure~\ref{fig:map}, with the filter icon to the left of the menu (gear) icon).
Filtering updates a bit vector in the original backing table, removing associated tuples from subsequent selection and reducing the opacity of corresponding marks to zero.

\textbf{Sort}.
Charts with categorical domains can be sorted in ascending or descending order by numeric data fields.
Users can move marks with numeric data extrema to an edge, generating a \textit{transform} predicate of the form (\texttt{field}: \texttt{NUMERIC\_FIELD}, \texttt{op}: \texttt{ORDERBY}, \texttt{value}: \texttt{[ASC / DESC]}).
An \texttt{orderby} query is applied to the data table; DIVI materializes ordering by swapping each ordered mark's view position with the the view position of the original mark at that index.

\textbf{Navigate}.
We modify D3's \texttt{zoom} API to orchestrate DOM-level pan / zoom transforms.
When a zoom event is fired, a transform object is passed with the scaling and translation transforms for each axis; the scale for pan events equals 1.
Each axis transform generates a \textit{transform} predicate of the form (\texttt{field}: \texttt{AXIS\_FIELD}, \texttt{op}: \texttt{TRANSFORMBY}, \texttt{value}: \texttt{TRANSFORM\_OBJECT}).
We use the transform object across all propagated views to translate, scale, and clip all non-legend SVG marks (including axes by rescaling the axis scale's domain) based on their global position within the SVG view.
To prevent scaling of a \texttt{path} element's visible stroke width when zooming, we set the \texttt{vector-effect} CSS attribute to \texttt{non-scaling-stroke}.
By default, a double-click will zoom by a factor of two, centered at the mouse position, if no panning or zooming has occurred. 
After a pan or zoom, a double-click will instead reset the view to its initial state.

\textbf{Annotate}.
Users can enable annotations with the annotate icon.
Clicking anywhere in the SVG displays a popup for users to provide the annotation label, which is then placed at the current mouse position (Figure~\ref{fig:map}).
We use the d3-annotations package~\cite{d3annotation} to construct the annotation and apply it to the SVG.

\subsubsection{Interactive Disambiguation}
\label{disambiguation}
Several interactions are inherently ambiguous, as the same input might correspond to multiple user intents.
Dragging may indicate either panning or brushing, while double-click may either zoom or reset the selection state.
To enable disambiguation, upon input DIVI displays a toolbar \includegraphics[height=\fontcharht\font`\B]{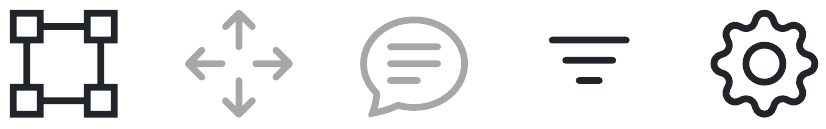} (Fig.~\ref{fig:map}(D)) with which users can adjust the interpretation of their input .
The toolbar overlays the SVG and is visible when the user's mouse enters the view and hidden when the mouse leaves the view.
Icons are initially displayed for brush, pan, and annotate interactions, with brushing inputs (drag and double click) enabled by default.
Clicking one icon disables the others and binds the corresponding interaction handlers to prevent input ambiguity.
When marks are selected, a filter icon appears.

\begin{figure*}[!t]
    \centering
    \resizebox{\linewidth}{!}{\includegraphics{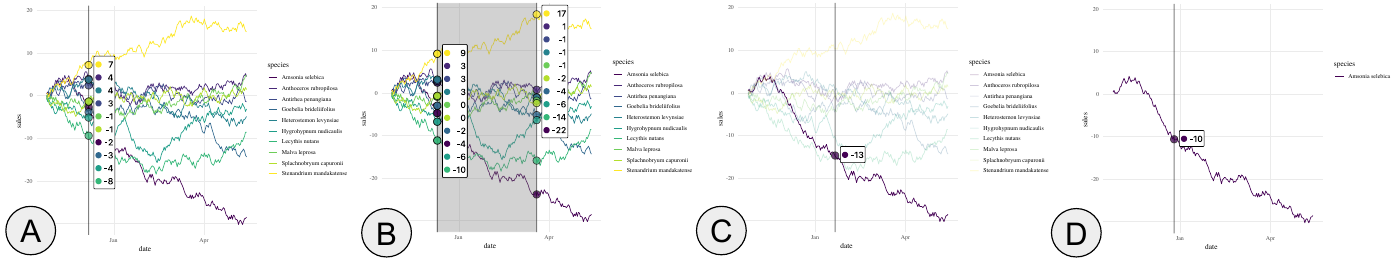}}
    \vspace{-15pt}
    \caption{DIVI applied to a ggplot2 multi-line chart. (A) Hover tooltips show each line's value. (B) Brushing displays tooltips for the start and end points. (C) Clicking a line selects it and limits the tooltip to just that line. (D) Inclusive filtering removes other lines and their legend keys.}
    \label{fig:ggplot}
    \vspace{-3mm}
\end{figure*}


\section{\textcolor{red}{SVG Inference Limitations}}
\label{limitations}

DIVI is applicable across a wide range of chart types, interactions, and source tools (\S\ref{eval:gallery}).
\textcolor{red}{
Although SVG output is supported by most visualization tools and many interactions can be realized directly as modifications to the SVG, SVG source content alone results in certain limitations as to what we can theoretically infer.
}

\begin{figure}[!b]
    \centering
    \begin{minipage}{.4\linewidth}
        \centering
        \includegraphics[width=\textwidth]{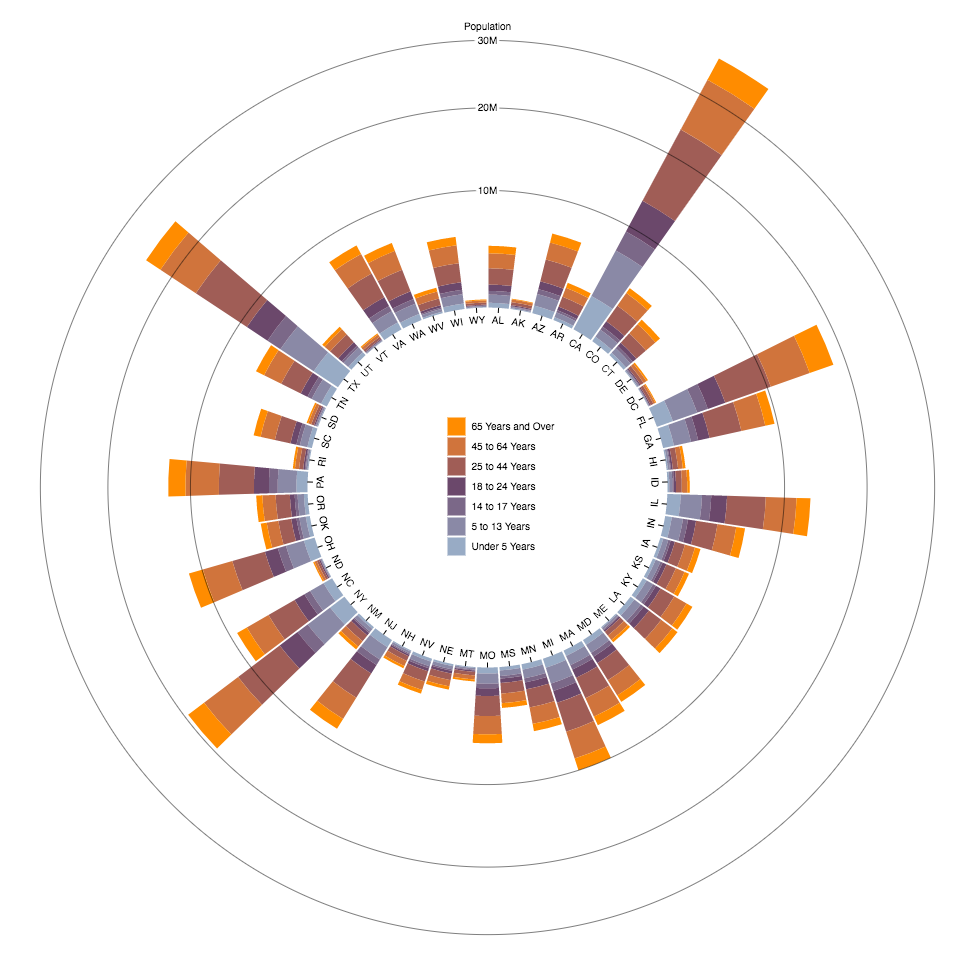}
    \end{minipage}%
    \begin{minipage}{.6\linewidth}
        \centering
        \includegraphics[width=\textwidth]{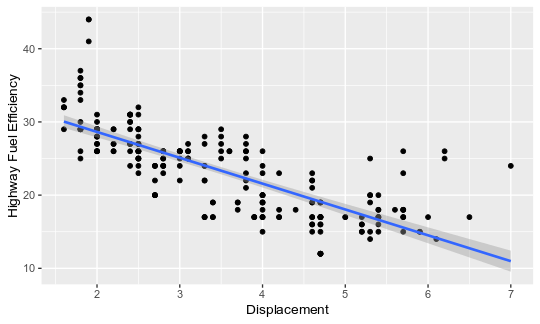}
    \end{minipage}
    \vspace{-10pt}
    \caption{Limitations of SVG inference: (Left) D3~\cite{bostock2011d3} stacked bar chart with circular $x$-axis. (Right) ggplot2~\cite{ggplot2} scatter plot with regression line and $\pm1\sigma$ bounds.}
    \label{fig:limitations}
    \vspace{-1mm}
\end{figure}

\textbf{Custom Charts \& Styling}.
DIVI's deconstruction process (\S\ref{deconstruction}) relies on positional and styling relations within chart structures (e.g., axes and legends) typical of many plotting tools.
However, highly custom graphics may be more difficult to deconstruct.
For example, circular axes (Fig.~\ref{fig:limitations}, left) are not aligned and, as a result, cannot be inferred. 
Dual axes also cannot be inferred \textcolor{red}{without color-coding}.
While users may still interact with individual SVG marks, interactions are not grounded in data space \textcolor{red}{(c.f., VisDock~\cite{choi2015visdock} view manipulation)}.

\textbf{Semantic Mark Structures}.
\textcolor{red}{
We tested DIVI with the following common chart types represented in our gallery (\S\ref{eval:gallery}): bar chart, scatter plot, (stacked) area chart, (multi) line chart, and geographic map.}
\textcolor{red}{However, since DIVI infers structures bottom-up instead of top-down to enable decoupling (DG2), global} abstractions typical of many statistical plots, \textcolor{red}{including network diagrams, iso-contours, and density plots}, may be impossible to infer from SVG alone.
\textcolor{red}{For instance, do circles with connected lines represent a line chart or network visualization?}
As another example, Fig.~\ref{fig:limitations} (right) displays a regression line with $\pm1\sigma$ bounds that may use external parameters (e.g., regularization) which cannot be inferred without external knowledge.
Custom handling would resolve this ambiguity; for instance, re-computing a regression curve based on selected scatter plot points.
Additional examples include box plots (which may comprise disparate \texttt{rect}, \texttt{line}, \texttt{path} SVG elements that cannot be grouped), continuous distributions (e.g., Kernel Desnity Estimation), parameterized marks such as regularized or polynomial regression curves, and trigonometric curves.

\textbf{SVG Element Representations}.
\textit{How} an individual mark is represented as an SVG element may also limit inference.
For instance, continuous color legends are often rendered as bitmap images rather than \texttt{linearGradient} elements, requiring additional inference of pixel-color mappings.
As another example, maps may not render regions as separate path elements; e.g., U.S. states may be generated as one large path element.
In Figure~\ref{fig:map} we specifically use different state topologies to enable interactive brushing.

\textbf{Custom Data Transformations}.
DIVI's linking inference is limited to \texttt{groupby}, \texttt{aggregate}, and binning and formatting transforms.
While certain transform and projection combinations can be inferred (e.g., \texttt{groupby}, \texttt{aggregate}, then \texttt{filter}), \textit{intermediate} transforms cannot.
For instance, a data table may be filtered prior to aggregating; however, DIVI cannot infer this without exhaustively enumerating all row combinations, aggregating them, and comparing the result---an infeasible computation for a reasonably-sized dataset.
Other examples include custom derivations (e.g., squared column values, standardization, normalization), formatting (e.g., abbreviation), and sampling.
Unless these derivations are reified in new data tables prior to additional projection or transformation, DIVI cannot link them to other data.

\section{Example Gallery}
\label{eval:gallery}

\textcolor{red}{To demonstrate DIVI's support across different interactions (DG1), chart types, linkings, and source tools (DG2), we produced a gallery of examples (Figures \ref{fig:teaser}--\ref{fig:ggplot}) that automatically add interaction to static visualizations (DG3) using inferred chart semantics (DG4)}.
Each example starts with one or more static SVGs produced from diverse tools (DG2) such as D3~\cite{bostock2011d3}, Vega-Lite~\cite{satyanarayan2016vega}, Matplotlib~\cite{matplotlib} gpplot2~\cite{ggplot2}, and Excel.
The examples include various mark types (points, bars, lines, areas, polygons for geographic regions), encoding channels (position, stacking, color, and size), chart configurations (single chart, multi-view, and cross-tool), and multi-view linkings (projection, filtering, and aggregation).
The examples also span the interaction types (selection, filter, sort, pan, zoom, etc.) and target elements (marks, legend, background, axes) identified in our taxonomy (Table~\ref{table:taxonomy}).

\textcolor{red}{
\section{Evaluation: Usability Study}
\label{usability_study}

We conducted a first-use study in which participants completed a series of interactive tasks drawn from our taxonomy.
We sought to assess how easily participants could use DIVI across a variety of interaction techniques and charts (DG1) and disambiguate their intents (DG3).

The 13 participants were graduate (4) or undergraduate (9) students, all but one studied computer science.
The average self-reported familiarity with data visualization was 3.0 ($\sigma = 0.39$) on a 5-point Likert scale.
The average experience with \textit{creating} visualizations was 2.8 ($\sigma = 0.58$). 
Participants reported using the following visualization tools: Matplotlib/Seaborn (7), D3 (7), Tableau (6), Vega-Lite/Altair (3), Observable Plot (3), Chart.js/p5.js (2), Excel (2), and R/ggplot2 (2).

\subsection{Methods}

Participants completed 22 tasks across multiple interaction and visualization types, using a web-based DIVI prototype.
For each task, participants were asked to perform one or more interactions (e.g., zoom, filter, select, etc.) using their mouse/trackpad and keyboard.
We asked participants to perform specific tasks to standardize the experience and ensure interaction coverage (DG1).
We did not guide participants as to \textit{how} the interaction should be performed, nor did we provide a DIVI tutorial, as the primary goal was to assess how easily users could perform and disambiguate the requested interactions.
Study sessions lasted 30 minutes and were conducted over Zoom video conferencing.
Participants were compensated with a \$15 gift card.

\subsection{Results}
\label{eval:results}

Participants completed 19/22 of tasks within $\sim$30 seconds (Table~\ref{table:tasks} in Appendix~\ref{study-results}), using input sequences they were familiar with.
Three of the tasks proved more challenging: (1) select all points except an outlier, (2) $x$-axis-constrained zooming, and (3) 1D-constrained brushing to select marks in a certain range. 
The outlier-sensitive selection task was difficult because users could not append new brush selections to prior selections (e.g., when pressing a meta-key); we added this feature mid-study and the issue resolved.
The other slower tasks ($x$-axis zooming, constrained brushing) suffered from discoverability issues: participants did not know they could interact with an axis to constrain navigation and selection.
Unlike standard 2D navigation and selection, participants were unsure how to convey constraints.
In response to these observations, we now update the pointer cursor (e.g., to left-right or up-down arrows) along axes to make the interactions more discoverable.
We also now include 1D constraints as control menu settings (Fig.~\ref{fig:map}(d)).


Participants used the toolbar to disambiguate their intents with ease. 
Reactions to DIVI were enthusiastic:
``\textit{really cool, robust toolset!}'',
``\textit{really helpful}'', 
``\textit{nice to look at [an] image and interact with it.}'' 
Several participants appreciated the advantages over static plots and lower-level libraries (``\textit{nice to play with a graph I don't have to recompile.}''), whereas creating interactions with other tools can be ``\textit{really hard}'' and ``\textit{difficult to implement.}''
One participant shared that they found the technical overhead of D3 to be too prohibitive for simple interactions such as range-based selections (e.g., brushing).


Participants disagreed regarding how difficult DIVI might be for users lacking prior visualization experience. 
Some felt that experience with graphical file managers could be sufficient
(e.g., users can apply file interface interactions from desktop computers to selection and brushing interactions in DIVI, also echoing studies of touch gestures~\cite{wobbrock2009user}).  
One participant (who struggled with x-axis brushing) remarked that they disliked the automatic features and would prefer a menu-based toolbar.
The other 12 participants instead preferred the DIVI approach.
To support varied user preferences, DIVI provides a control menu to configure the provided interactions (Fig.~\ref{fig:map}(D)).
}
\section{Discussion \& Future Work}
\label{sec:discussion}

We presented Dynamically Interactive Visualization (DIVI), a novel approach for decoupling interaction from chart specification to dynamically add interactions to static visualizations.
\textcolor{red}{
Motivated by interviews with six data analysts, DIVI contributes a three-part model to enable dynamic interaction: (1) SVG deconstruction to automatically infer requisite chart metadata (DG4); (2) standardized chart representation to decouple specification and interaction logic (DG2); and (3) automatic interaction handling (DG3).
This approach leverages an interaction taxonomy that categorizes user interactions by target visual element, interaction type, and input event, and outlines what chart metadata is needed to achieve each interaction.
We analyze static SVG images to automatically extract required metadata: axes, legends, and data-representative marks; infer multi-view linkings; and coordinate user input on SVG elements.
We provide an example gallery and usability study with 13 participants demonstrating that DIVI can support an expressive gamut of interactions (DG1), chart types, and source tools (DG2), while enabling users to easily express and disambiguate their interaction intents (DG3).
}

Going forward, we plan to extend DIVI to support touch input on mobile devices.
We believe the same general approach of interaction decoupling---dynamically deconstructing charts and coordinating user interactions---can be applied, though this awaits evaluation.
Another important avenue concerns accessibility, including keyboard navigation (e.g., via tab and arrow keys).
Our SVG analysis and selection interactions might be used to automatically add at least a rudimentary level of screen reader compatibility.

\textcolor{red}{
In the future, we plan to supplement DIVI with custom handlers, specification options, or ML-driven disambiguation that users may engage with to resolve the potential inference failures (\S\ref{limitations}).
If dynamic interaction proves valuable in practice, we envision augmenting visualization tools to also output specification metadata, such as scale mappings or data bindings, to ease inference and support a wider array of interactions.
Standardizing such metadata across tools could enable richer interoperability.
Meanwhile, computer vision techniques for reverse-engineering bitmap image visualizations~\cite{savva2011revision, poco2017reverse, poco2017extracting} might expand the scope of DIVI further.
In addition, while DIVI supports faceted linking, certain tools nest faceted plots within the same SVG tree, requiring additional inference to extract them in the future.
}

So far we have focused on exploratory visualization.
However, we believe DIVI also provides a promising direction for enhancing storytelling and presentation, as interactions needed for annotation or recording purposes would no longer need to be manually programmed in advance.
Authoring and annotation environments leveraging DIVI could make it easier to construct dynamic interactive visualizations from static views, and let viewers annotate interesting insights for later use or sharing.
Further, DIVI could be used to construct narrative sequences, introducing a dataset (e.g., showing axes first, then adding data layers) and cycling through annotated selections.
Extending DIVI with ``record'' and ``playback'' options could support dynamic figure authoring and improve communication.

Our implementation of DIVI is available as open-source software at \url{https://uwdata.github.io/DIVI/}.

\acknowledgments{
We thank Leilani Battle and the UW Interactive Data Lab for their valuable feedback. 
This work was supported by a Moore Foundation software grant.
}

\bibliographystyle{abbrv-doi-hyperref}

\bibliography{bibliography}

\appendix

\section{SVG Deconstruction Algorithms}
\subsection{Initial Text \& Mark Grouping}
\label{appendix:grouping}
DIVI exploits positional relations (alignment, style grouping, and distance) between text and SVG marks to infer chart components. 
Text and marks are separately grouped by alignment first.
DIVI uses the DOM \texttt{getBoundingClientRect()} API and its \texttt{left}, \texttt{right}, \texttt{top}, and \texttt{bottom} fields to group each element into alignment clusters.
Pairs of text- and mark-aligned clusters are then examined to disambiguate axis, legend, and mark elements.

\subsection{Axes}
\label{appendix:axes}
DIVI splits each mark-aligned cluster into styled groups; groups above a maximum distance threshold from a candidate text group (set at $10\%$ of the viewport width or height, whichever is larger) are pruned to reduce the search space.
Text and mark groups are finally sorted in their non-aligned direction (e.g., if vertically-aligned, they will be sorted by their $x$ position). Element-wise pairings with minimized centered distance offsets are flagged as ticks comprising (\texttt{label}, \texttt{mark}) tuples.


\subsection{Legends}
\label{appendix:legends}
DIVI prunes candidate mark groups without at least one differing style field (color, size, or shape).
As with axes, groups above the maximum distance threshold from a candidate text group ($\%10$ of viewport \texttt{width} or \texttt{height}) are removed, and element-wise mark / text pairings are minimized to identify legends.
Remaining (unpaired) text marks outside of the axes (to the left / right of the horizontally-aligned $y$ axis or to the bottom / top of the vertically-aligned $x$ axis) are identified as axis or chart titles based on closest distance.

\subsection{Scales}
\label{appendix:scales}
Tick labels from inferred axes are parsed as numeric, string (category), or date types.
For numeric and date labels, we use the extreme (min / max) values as the scale domain.
For instance, if the $x$-axis domain is $[0, 100]$, the leftmost tick label will be 0 and the rightmost tick label will be 100.
To support log scales, we consider two possibilities: Are the tick marks or tick label values non-linearly spaced?
If either holds, we assume a log scale. 
We use the view placement of the ticks (e.g., centered $x$ position for the $x$-axis) to compute the scale range (i.e., pixel placement within SVG view).
If an axis was not identified ($x$, $y$, or both), we use respective SVG \texttt{width} and \texttt{height} attributes as the scale domains (i.e., the scale mapping is the identity function).

\subsection{Views}
\label{appendix:views}
DIVI infers stacking groups from $x$ or $y$ position alignment with even spacing.
Numeric linear scales (linear, log, or date) for stacking groups (bar chart, dot plot, etc.) are inferred as histogram bin ranges between sequential tick pairs.
To detect vertical or horizontal chart orientation, DIVI analyzes their $x$ and $y$ positions: Vertical charts share the same $y$ position, whereas horizontal charts share the same $x$ position.
To account for diverging bar directions, DIVI inspects shared \texttt{top} or \texttt{bottom} $y$ positions for vertical orientation (or \texttt{left} and \texttt{right} for horizontal orientation) to identify the starting axis.

\subsection{Data Analysis}
\label{appendix:data-analysis}
For each discrete mark or point in a continuous mark (line or area), DIVI maps the mark's (1) position through inverted $x$- and $y$-axis scales and (2) style encodings through each legend scale to reconstruct data fields.
Inferred data for each mark are merged into a single relational table.
To support efficient indexing, marks maintain their corresponding table row indices, while each row maintains the corresponding source mark.
We use Arquero~\cite{arquero} for query processing.

\section{Usability Study Results}
\label{study-results}
\begin{table}[!ht]
\centering
\setlength\tabcolsep{0pt} 
\smallskip 
\scalebox{0.73}{
\begin{tabular*}{\linewidth}{@{\extracolsep{\fill}} cccc}
\toprule
\multirow{1}{*}{Visualization} & 
\multirow{1}{*}{Task} & 
Avg Time (StdDev) \\
\toprule 
  \multirow{8}{*}{Scatter plot}  & Zoom $x$-axis & 1:56 (1:16) \\
  & Zoom $y$-axis & 0:18 (0:09) \\
  & Brush points & 0:16 (0:12) \\
  & Zoom/pan to fit & 0:19 (0:16) \\
  & Filter points & 0:07 (0:08) \\
  & Tooltip & 0:18 (0:16) \\
  & Select points & 1:03 (0:53) \\
  & Annotate & 0:17 (0:12) \\
  
  \midrule
  \multirow{2}{*}{Line chart}  & Tooltip & 0:28 (0:28) \\
  & Brush & 0:32 (0:23) \\
   
  \midrule
    \multirow{3}{*}{Stacked area chart} & Brush  & 0:27 (0:16) \\
     & Select and filter & 0:25 (0:17) \\
     & Unfilter & 0:02 (0:02) \\
  
  \midrule
     \multirow{3}{*}{Bar chart} & Sort & 0:25 (0:34) \\
     & Brush & 0:18 (0:11) \\
     & Filter & 0:04 (0:09) \\
     
  \midrule
     \multirow{1}{*}{Log line chart} & Brush & 0:15 (0:16) \\
     
  \midrule
     \multirow{2}{*}{Hexbin chart} & Tooltip & 0:07 (0:07) \\
     & Select and filter & 0:10 (0:08) \\
  
  \midrule
     \multirow{2}{*}{Stacked bar chart} & Legend selection & 0:27 (0:29) \\
     & Brush and filter & 0:24 (0:17) \\
  
  \midrule
     \multirow{1}{*}{Scatter plot} & 1D brush and filter & 0:53 (0:35) \\
     
\bottomrule
\end{tabular*}
}
\vspace*{1mm}
\caption{Study results for 22 sequential visualization tasks, with average and standard deviation completion times in minutes.
}
\label{table:tasks}
\end{table}

\newpage
\section{Linking Algorithm}
\label{linking-algorithm}
Here we detail the algorithm for linkage inference across multi-view SVG visualizations.
\SetKwComment{Comment}{/* }{ */}

\begin{algorithm}[!ht]
\caption{Linking Algorithm}
\label{alg:two}
\SetKwFunction{FMain}{RL}
\SetKwProg{Fn}{Function}{:}{}
\Fn{\FMain{$D_1, D_2$}}{
    \If{$\texttt{numRows}(D_1) > \texttt{numRows}(D_2)$}{
        \textbf{return} \texttt{``NONE''}\;
    }
    $M, F \gets \{\}$\;
    \For{$a_i \in D_1$}{
        $D_1[a_i] \gets \texttt{sort(}D_1[a_i]\texttt{)}$\;
        \For{$b_j \in D_2$} {
            \If{$F[b_j]$}{\textbf{continue}\;}
            $D_2[b_j] \gets \texttt{sort(}D_2[b_j]\texttt{)}$\;
            \If{$D_1[a_i] \subseteq D_2[b_j]$}{
                $M[a_i] \gets b_j$\;
                $D_2 \gets \text{all } \textbf{rows}(D_2)$ 
                \text{(including duplicates)} \text{where matched}\;
                $F[b_j] = \texttt{true}$\;
                \textbf{break}\;
            }
        }
    }

    \textbf{return} $M, D_2$\;
}
\textbf{End Function}

\vspace{4pt}

\SetKwFunction{FMain}{TL}
\SetKwProg{Fn}{Function}{:}{}
\Fn{\FMain{$D_1$}}{
    \For{$a_i \in D_1$} {
        \If{$a_i$ \text{is transformed}}{
            \textbf{continue}\;
        }
        \For{$t_j \in T$}{
            $D' \gets D_1$\;
            $D'[a_i] \gets t_j(D_1[a_i])$\;
            \textbf{yield} $D'$\;   
        }
    }
}
\textbf{End Function}

\vspace{4pt}

\SetKwFunction{FMain}{getLink}
\SetKwProg{Fn}{Function}{:}{}
\Fn{\FMain{$D_1, D_2$}}{
    $M, D \gets \texttt{RL}(D_1, D_2)$\;
    \If{$M != \texttt{``NONE''} \wedge !\texttt{empty}(M)$}
    {\textbf{return} $M, D$\;}
    \For{$D'\text{of \texttt{TL}}(D_1)$} {
        $M, D \gets \texttt{getLink}(D', D_2)$\;
        \If{$M != \texttt{``NONE''}$ $\wedge$ $!\texttt{empty}(M)$}
        {\textbf{return} $M, D$\;}
    }
    \textbf{return} \texttt{``NONE''}\;
}
\textbf{End Function}

\vspace{4pt}

\SetKwFunction{FMain}{Linking}
\SetKwProg{Fn}{Function}{:}{}
\Fn{\FMain{$D_1, D_2$}}{
    $L \gets \texttt{getLink(}D_1, D_2\texttt{)}$\;
     \If{$ L = \texttt{``NONE''} $}{
        {$L \gets \texttt{getLink(}D_2, D_1\texttt{)}$}\;
        }
        \textbf{return} $L$\;
}
\textbf{End Function}
\vspace{4pt}
\end{algorithm}

\end{document}